# Spatiotemporal Analysis of Ridesourcing and Taxi Demand by Taxi-zones in New York City

**Patrick Toman**
Department of Statistics
University of Connecticut
Philip E. Austin Building Rm 341, Storrs, CT 06269
Tel: 303-710-1209 Email: patrick.toman@uconn.edu

**Jingyue Zhang**
Department of Civil and Environmental Engineering
University of Connecticut
Castleman Building, Storrs, CT 06269
Tel: 860-208-8591; Email: jingyue.zhang@uconn.edu

**Nalini Ravishanker**
Department of Statistics
University of Connecticut
Philip E. Austin Building Rm 333, Storrs, CT 06269
Tel: 860-486-4760; Email: nalini.ravishanker@uconn.edu

**Karthik C. Konduri**
Department of Civil and Environmental Engineering
University of Connecticut
Castleman Building Rm 331, Storrs, CT 06269
Tel: 860-486-2733; Email: karthik.konduri@uconn.edu

# Abstract

The burst of demand for TNCs has significantly changed the transportation landscape and dramatically disrupted the Vehicle for Hire (VFH) market that used to be dominated by taxicabs for many years. Since first being introduced by Uber in 2009, ridesourcing services have rapidly penetrated the market. This paper aims to investigate temporal and spatial patterns in taxi and TNC usage based on data at the taxi zone level in New York City. Specifically, we fit suitable time series models to estimate the temporal patterns. Next, we filter out the temporal effects and investigate spatial dependence in the residuals using global and local Moran's I statistics. We discuss the relation between the spatial correlations and the demographic and land use effects at the taxi zone level. Estimating and removing these effects via a multiple linear regression (MLR) model and recomputing the Moran's I statistics on the resulting residuals enables us to investigate spatial dependence after accounting for these effects. Our analysis indicates interesting patterns in spatial correlations between taxi zone sin NYC and over time, indicating that predictive modeling of ridesourcing usage must incorporate both temporal and spatial dependence.

**Keywords**: Moran's I, ridesourcing, spatial clustering, taxi zones in NYC, temporal dependence.

# 1. Introduction

Ridesourcing, also known as Transportation Network Company (TNC), like Uber, Lyft, or Via, provide mobility services by allowing passengers to use their smartphone apps to connect with nearby drivers who typically drive part-time using their own car. The burst of demand in TNCs has significantly changed the transportation landscape and dramatically disrupted the Vehicle for Hire (VFH) market that used to be dominated by taxicabs for many years. Since first being introduced by Uber in 2009, ridesourcing services have rapidly penetrated the market. The total number of passengers transported by TNCs increased 37% from 1.90 billion in 2016 to 2.61 billion in 2017 (Schaller, 2018). In New York City, TNC ridership doubled annually from 2014 to 2016 and matched the ridership of Yellow Cabs in 2016 (Schaller, 2017). While February 2017 marks the first time that TNCs made more trips than yellow and green taxi combined, by December 2017, TNCs had made 65% more pickup trips than taxis in New York City (Schneider, 2018), most of these passengers being serviced by Uber or Lyft. The increasing adoption and popularity of TNCs raise a number of interesting questions. Supporters of TNCs envision ridesourcing as a fast, flexible, and convenient mobility service that can serve as an attractive alternative to driving and also filling gaps in the public transit network. They claim that TNCs can potentially reduce vehicle ownership and promote multi-modal usage in urban areas (Silver and Fischer-Baum, 2014). On the other hand, recent studies have found that TNCs take passengers away from public transit, increase vehicle miles traveled (VMT), and attenuate congestion (Erhardt et al., 2019; Qian et al., 2020).

Taxis and TNCs exhibit similar characteristics and provide comparable services, and there is an emerging body of scholarship on comparisons between them. Rayle et al. (2015) analyzed how TNCs are used in San Francisco by collecting 380 surveys in spring 2014. They found that ridesourcing passengers are typically young people who possess a higher education level compared to taxi users. Also, the survey data indicated that the TNCs not only replace taxis, they also replace public transit and driving. Schaller (2018) analyzed trip and user characteristics of TNCs in New York City. His report revealed similar results, i.e., TNCs riders are relatively young and well educated. He also pointed out that TNC trips are highly concentrated in large, dense urban areas, while the taxis generally tend to serve riders in suburban and rural areas. His report further indicated that TNCs are competing with public transit and nonmotorized modes such as walking and biking. Analyzing pickup data in New York City, *FiveThirtyEight* found both competing and supplementary relationships between TNCs and taxis. Their analysis indicated that Uber is replacing Yellow Cab rides in the center of the city and is supplementing Green Cab rides in the outer boroughs (Fischer-Baum and Bialik, 2015). Poulsen et al. (2016) compared ride records of Green Cabs and Uber in the outer boroughs of New York City at the zip code level, and showed that Green Cab and Uber rides exhibit similarities in terms of temporal trend and spatial distribution. Their results also confirmed that Green Cabs are losing market share to Uber. Gerte et al. (2019) analyzed the temporal trend of TNC demand in New York City and its impact on the existing ridesharing modes namely taxi, Subway and Citi Bike. Their results from fitting a dynamic linear model (DLM) to daily ridership usage via the different modes, reinforced a substitutionary relationship between TNCs and taxis and a complementary relationship between TNCs and Subway. These studies highlight the fact that despite the similarity in infrastructure, TNCs and taxis exhibit significant differences in user characteristics, trip purposes and spatial demand. The impact of TNCs on taxicabs needs to be carefully evaluated in light of the potential implications of any future policy changes in regulation and planning.

The primary purpose of this paper is to discuss spatial correlation in TNC and taxi usage at the taxi zone level in New York City. We first describe temporal modeling of ridesourcing data using suitable time series models and subsequent spatial analysis of resulting residuals from these models in order to understand relationships between the modes at this spatial resolution. Specifically, we fit autoregressive integrated moving average/Generalized autoregressive conditionally heteroscedastic (ARIMA-GARCH) models to weekly aggregated trip counts of TNC and Taxi in New York City from January 1, 2015 to June 30, 2017. We then analyzed the temporally filtered data (residuals from the time series models) to investigate spatial associations between taxi zones, after accounting for land use and demographic information at the taxi zone level.

The format of our paper is as follows. In Section 2, we discuss initial data processing and exploration. In Section 3, we describe temporal analysis via ARIMA-GARCH model fitting. This is followed in Section 4 by a regression of the time series model residuals on land use and demographic variables as predictors. Section 5 then describes spatial association analysis on the residuals from the multiple regression using global and local Moran's I statistics. Section 6 provides a discussion and summary of this stepwise procedure.

## 2. Data Processing and EDA

TNC and taxi trip data were obtained from New York City at the taxi zone level and span the time period from January 1st, 2015 until June 30th, 2017. The NYC Taxi and Limousine Commission (TLC) oversees all taxis and other types of for-hire vehicles including TNCs (TLC, 2017). TNC data was collected for the following ride modes: Uber, Lyft, and Via. Taxi data includes Yellow Cabs and Green Cabs. TLC made disaggregate "trip"/ "trip end" data available for TNC and Taxis. TNCs data contains precise coordinates of trip pickup location and timestamp for trip pickup time in 2014. However, after 2014, trip pickup location has been aggregated to predefined taxi zone level. Taxi data are available since 2009. The data also contains precise coordinates for where the trip started and ended and also the timestamp for when the trip started and ended. Gerte et al. (2019) preprocessed the data, aggregating daily demand across all five boroughs of NYC. They also merged in other sets of data such as those pertaining to weather events, holidays, and event permits. Gerte et al. (2019) and other researchers have found that there are three main types of exogenous variables that would potentially impact TNCs and other shared mode usage namely temporal, environmental and land use variables. In this article, a variety of temporal variables are explored to capture the temporal trend and seasonality in TNCs and other shared mode usage. In addition, The NYC Open Data portal was used to gather data about events and holidays (NYC Open Data, 2017). The data was primarily used to generate indicators as well as daily counts of event permits to be used as exogenous predictors in our analysis. Daily weather data comes from the National Oceanic and Atmospheric Administration (NOAA, 2017); are used to quantify how environmental factors influence TNCs and other shared mode usage. More specifically, we use the average of daily precipitation readings from three weather stations as exogenous predictors in our multivariate model.

Missing data were found in the Green Cab usage in 2016. The last couple of days of every month from July to December 2016 were missing, for a total of 24 missing days. To treat this, we employed a local averaging method where data from the same weekday were averaged and the missing value replaced. For example, suppose Green Cab data was missing for Tuesday, 2016-07-26. Observed Green cab data for every Tuesday were averaged and the missing value replaced by this average. Verification of this data was done by cross referencing yearly totals used in this paper with MTA's reported subway ridership and other published records. Our analysis is restricted to the time period spanning from 2015-01-11 to 2017-06-25.

There are 263 taxi zones in New York City; however, some of these zones (example, Governor's Island/Ellis Island/Liberty Island and Central Park) have very low daily trip counts (less than 10) for either TNC or Taxi. We aggregated the daily trip counts to weekly trip counts. The aggregated data consists of TNC trip counts for 229 zones and Taxi trip counts for 212 zones over 129 weeks starting from 2015-01-11 to 2017-06-25. The three exogenous variables, average precipitation (inches), number of city permitted events, and a dummy variable to indicate if any holiday is included in a week, were also aggregated to the weekly level. In most of the zones, TNC trip counts show a continuous increase over time, while the Taxi trip counts exhibit a decreasing trend. Figure 1 shows the weekly usage of TNC and Taxi in eight different zones in NYC. In each zone, TNC usage shows an increasing trend from 2015 to 2017, while the trend of Taxi usage varies between the zones. Taxi usage in Astoria, Queens and Bedford, Bronx show a clear decreasing trend from 2015 to 2017, while the usage in the other zones show some fluctuations rather than any large decrease.

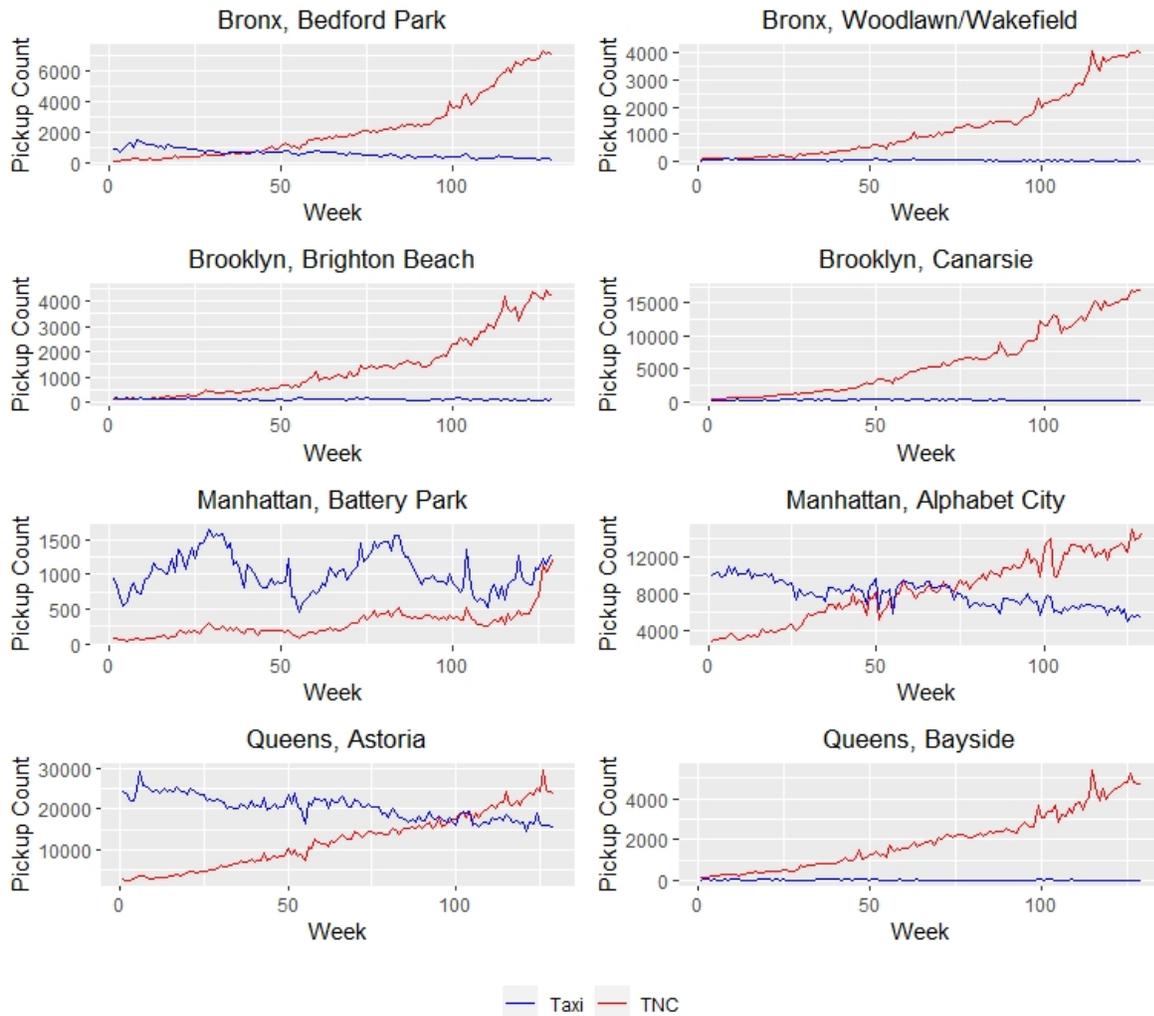

**Figure 1. Weekly TNC and Taxi Usage in Eight NYC Taxi Zones**

# 3 ARIMA-GARCH Models for Temporal Pattern Analysis

Let $X_t$ denote the usage in a single taxi zone. Initial exploratory analysis via the autocorrelation and partial autocorrelation functions suggest modeling the temporal patterns in $X_t$ by first fitting a suitable short-memory time series model, followed by accommodating the conditional heteroscedasticity by fitting a generalized conditional heteroscedastic (GARCH) model to the residuals. The multiplicative seasonal $ARIMA\,(p,d,q) \times (P,D,Q)_s$ model has the form (Shumway and Stoffer, 2017)

$$\boldsymbol{\phi}(B)\boldsymbol{\Phi}(B^s)\big((1-B)^d(1-B^s)^D X_t - \mu\big) = \boldsymbol{\theta}(B)\boldsymbol{\Theta}(B^s)W_t$$

where, $W_t \sim WN(0,\sigma^2)$; $\mu$ is the mean of $(1-B)^d(1-B^s)^D X_t$ and $\sigma^2$ is the white noise (WN) variance. Here, $d$ and $D$ are respectively the nonseasonal and seasonal (integer) degrees of differencing and $s$ is the seasonal period. $\phi(B) = 1 - \phi_1(B) - \cdots - \phi_p B^p$ is the nonseasonal AR polynomial of order $p$, $\theta(B) = 1 - \theta_1 B - \cdots - \theta_q B^q$ is the seasonal AR polynomial of order $q$ and $\Phi(B^s) = 1 - \Phi_1 B^s - \cdots - \Phi_P B^{sP}$ is the seasonal AR polynomial of order $P$ and $\Theta(B^s) = 1 - \Theta_1(B^s) - \cdots - \Theta_Q(B^{sQ})$ is the seasonal MA polynomial of order $Q$. In our data, we see that d = 1, D = 0, and s = 7 in all zones (see Table 1). We fit the seasonal ARIMA models to weekly TNC and Taxi usage to log $X_t$ in each taxi zone by using the auto.arima function in R. Table 1 summarizes the seasonal ARIMA results for both TNC and Taxi usage. Let $r_t$ denote the residuals from fitting the model above to $X_t$. The Ljung-Box and McLeod-Li tests (Shumway and Stoffer, 2017) enable us to evaluate whether the residuals are and the squared residuals are uncorrelated zero mean white noise sequences. Squared residuals failing the McLeod-Li test indicate the presence of conditional heteroscedasticity. This information is summarized in Table 2. If there is evidence of conditional heteroscedasticity, we may fit a $GARCH(m,s)$ model to $r_t$

$$r_t = \sigma_t \varepsilon_t$$
$$\sigma_t^2 = \alpha_0 + \sum_{j=1}^{m} \alpha_j r_{t-j}^2 + \sum_{j=1}^{s} \beta_j \sigma_{t-j}^2$$

where $\varepsilon_t \sim i.i.d.\,N(0,1)$. To ensure positivity of $\sigma_t^2$ and model stationarity, we assume that $\alpha_0 > 0$, $\alpha_i \geq 0$ for $i = 1, \cdots, s$ and $\sum_{i=1}^{\max(m,s)} (\alpha_i + \beta_j) < 1$.

Our analysis showed that 91 (of the 229) zones corresponding to TNC usage and 72 (of the 212) zones for Taxi usage required a GARCH (1, 1) model fit to the residuals from the seasonal ARIMA model. The resulting residuals were temporally uncorrelated, except in a few zones both for TNC and Taxi, although the correlation was not very high, see Table 2.

Figure 2 shows maps of these taxi zones for both modes, i.e., TNC and Taxi. The white spots on the maps are the zones excluded from our further analysis since their residuals fail either the Ljung-Box or McLeod-Li tests. The borough color coded zones are the temporally uncorrelated zones that we used for the exploration of spatial associations in TNC and taxi usage. Future analysis will expand the scope of the temporal modeling to include all taxi zones.

Table 1. Seasonal ARIMA Model Orders by Taxi Zones

| TNC | | Taxi | |
|---|---|---|---|
| ARIMA Model $(p,d,q) \times (P,D,Q)_s$ | Number of Zones | ARIMA Model $(p,d,q) \times (P,D,Q)_s$ | Number of Zones |
| $(1,1,0) \times (0,0,0)_4$ | 9 | $(1,1,0) \times (0,0,0)_4$ | 6 |
| $(1,1,1) \times (0,0,0)_4$ | 33 | $(1,1,1) \times (0,0,0)_4$ | 13 |
| $(1,1,1) \times (0,0,1)_4$ | 6 | $(1,1,1) \times (0,0,1)_4$ | 7 |
| $(1,1,1) \times (1,0,0)_4$ | 7 | $(1,1,1) \times (1,0,0)_4$ | 16 |
| $(1,1,3) \times (0,0,0)_4$ | 4 | $(3,1,0) \times (0,0,0)_4$ | 5 |
| $(2,1,1) \times (0,0,0)_4$ | 15 | $(0,1,1) \times (0,0,0)_4$ | 66 |
| $(2,1,2) \times (0,0,0)_4$ | 3 | $(0,1,1) \times (0,0,1)_4$ | 16 |
| $(0,1,1) \times (0,0,0)_4$ | 69 | $(0,1,1) \times (1,0,0)_4$ | 6 |
| $(0,1,1) \times (0,0,1)_4$ | 7 | $(0,1,1) \times (1,0,1)_4$ | 6 |
| $(0,1,1) \times (0,0,2)_4$ | 4 | $(0,1,2) \times (0,0,0)_4$ | 19 |
| $(0,1,1) \times (1,0,0)_4$ | 8 | $(0,1,2) \times (1,0,0)_4$ | 11 |
| $(0,1,2) \times (0,0,0)_4$ | 25 | $(0,1,3) \times (0,0,0)_4$ | 5 |
| Other | 39 | Other | 36 |

Table 2. Summaries from Ljung-Box and McLeod-Li Tests for ARIMA-GARCH Model Residuals

| Mode | Number of Taxi Zones Fail Ljung-Box Test (p<0.05 and Lag =12) | Number of Taxi Zones Fail McLeod-Li Test (p<0.05 and Lag =12) | Total Number of Temporally Uncorrelated Zones | Total Number of Taxi Zones |
|---|---|---|---|---|
| TNC | 16 | 2 | 213 | 229 |
| Taxi | 14 | 2 | 198 | 212 |

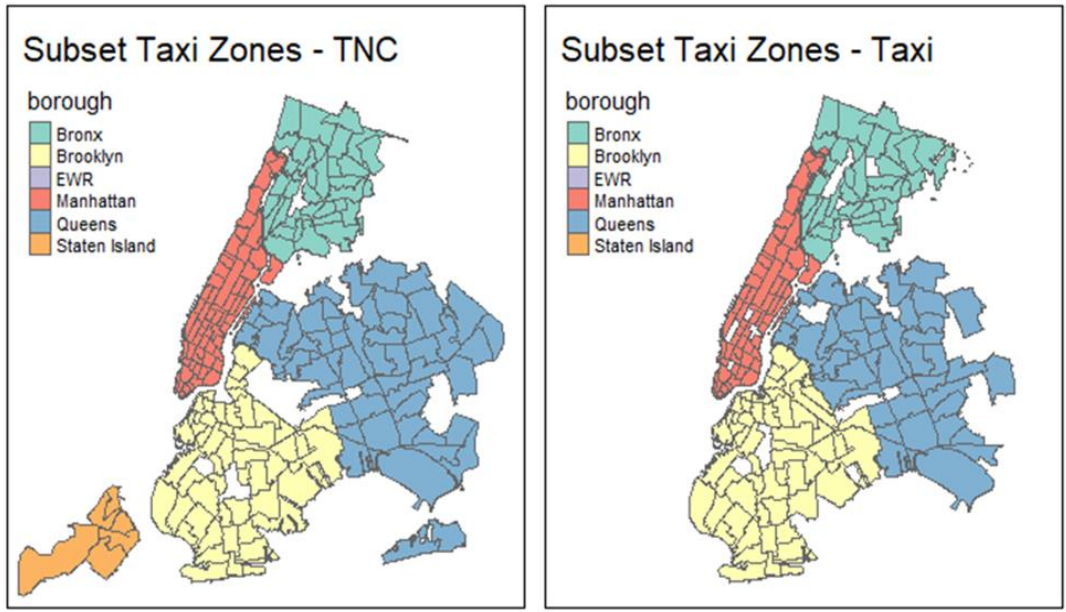

Figure 2. Taxi Zones with Temporally Uncorrelated Residuals for TNC and Taxi Usage

# 4. Multiple Linear Regression of ARIMA-GARCH Residuals on Land Use and Demographic Variables

After estimating and filtering out temporal patterns from both TNC and Taxi usage, we fit a multiple linear regression (MLR) model to the ARIMA-GARCH residuals in order to estimate potential influence of land use and demographic characteristics. This section presents a short description of census and land use data used for the MLR modeling and summarizes the estimation results.

## 4.1 Census and Land Use Data

The U.S. Census Bureau annually collects a variety of demographic information about the U.S. population and economy at different geographical resolutions. For our analysis, we focus on the total population, number of full time employees, median age and median earning in each taxi zone in New York City during the time period of our investigation. The data are downloaded from the U.S. Census Bureau website. Note that the data is reported at census track levels which are smaller than taxi zones. An aggregation is thus required to obtain these data at the taxi zone level. Land use data for New York City is collected by the Department of City Planning (DCP) which is a primary land use agency in New York City. DCP collects detailed land use and geographic data annually at the tax lot level including residential area, commercial area, retail area, etc. The dataset also provides the census track id for each tax lot. Therefore, we performed a similar aggregation to obtain the land use data for each taxi zone from 2015 to 2017.

Let Total Area be the sum of the lot area and the building area. We then construct proportions of Total Area for (i) Residential Area, (ii) Commercial Area, (iii) Retail Area, (iv) Factory Area, (v) Storage Area, (vi) Garage Area, (vii) Office Area, and (viii) Other Area, in order to obtain eight covariates pertaining to land use in each taxi zone. We used the following four demographic variables: (a) Total Population/# of Buildings, (b) Fulltime Employed/# of Buildings, (c) Median Age, and (d) Median Earnings.

## 4.2 MLR Model Results

Let $e$ denote an n-dimensional vector of ARIMA-GARCH residuals from a taxi zone, and let $X$ denote the matrix of regressors consisting of all main effects for the demographic and land use variables together with all second-order interactions between these variables. The MLR model is

$$e = X\beta + \eta$$

where, $\beta$ is the vector of partial regression coefficients and the error vector $\eta \sim N(0, \sigma^2 I_n)$ where $I_n$ denotes the nxn identity matrix. Stepwise model selection is performed using the MASS library in R (Ripley et al., 2019). The selection criterion is AIC and both forward and backward selections are utilized.

Two different regression analysis were built for each mode (TNC and Taxi). The first model incorporates only the main effects while the second model includes all main effects plus the second order interactions. For both analyses, we used stepwise model selection based on the AIC in order to find a parsimonious subset of the predictors. Regression output for these models are given below. Note that stepwise selection gave the same model for Taxi in both cases. We use the 5% level of significance. Stepwise model selection reveals that for Taxi, only the number of fulltime employed individuals has a statistically significant correlation with the response, see Table 3.

**Table 3 MLR Results for Taxi**

| Taxi-Best Model According to AIC | | | | |
|---|---|---|---|---|
|  | Term | Std. Error | T-Statistic | P-Value |
| Intercept | -0.003 | 0.001 | -2.521 | 0.012 |
| Fulltime Employment | -1.000e-08 | 4.336e-09 | -2.307 | 0.022 |
| Observations | 25413 | | | |
| Adjusted $R^2$ | 0.0002094 | | | |
| Residual Std. Error | 0.1491 | | | |
| F-Stat (df=1;25411) | 5.5322 | | | |

For TNC, a larger subset of covariates were statistically significant at the 5% level. For the main effects, full time employment, median earnings, and median age had statistically significant relationships with the response. In addition, interaction effects between the residential land use percentage and median age, office land use percentage and median age, and commercial percentage and fulltime employment were significant, see Table 4.

**Table 4 MLR Results for TNC**

| TNC-Best Model According to AIC | | | | |
|---|---|---|---|---|
|  | Term | Std. Error | T-Statistic | P-Value |
| Intercept | -0.005 | 0.004 | -1.3405 | 0.180 |
| ResidentialPct | 0.024 | 0.013 | 1.778 | 0.075 |
| FulltimeEmp | -2.311e-08 | 5.487e-09 | -4.213 | 2.53e-05 |
| Median Earnings | -2.151 | 8.391 | -2.564 | 0.010 |
| Median Age | 0.001 | 0.0001 | 3.467 | 0.0005 |
| ResidentialPct:Median Earnings | 3.418e-07 | 1.900e-07 | 1.828 | 0.068 |
| StoragePct:Median Earnings | 1.022e-06 | 6.949e-07 | 1.472 | 0.141 |
| ResidentialPct:Median Age | -0.0009 | 0.0004 | -2.097 | 0.0360 |
| OfficePct:MedianAge | 0.0006 | 0.0002 | 2.950 | 0.0032 |
| CommercialPct:FulltimeEmp | -9.396e-08 | 2.949e-08 | -3.187 | 0.0014 |
| Observations | 27477 | | | |
| Adjusted $R^2$ | 0.005 | | | |
| Residual Std. Error | 0.1234 | | | |
| F-Stat (df=9;27467) | 15.563 | | | |

## 5. Spatial Analysis of ARIMA-GARCH Residuals

We study spatial associations between the residuals from the different taxi zones after fitting and removing temporal dependence and adjusting for demographic and land use effects. Specifically, we use Moran's I to quantity the association of a zone with its neighborhood. Both global and local Moran's I values are calculated to investigate the overall and local spatial correlation, using the R package *spdep* (Bivand, 2018), see https://cran.rproject.org/web/packages/spdep/spdep.pdf

## 5.1 Global Moran's I

Moran's I is a statistic which computes the cross-product between a variable and its spatial lag, where the variable is expressed in deviations from its mean. The global Moran's I is calculated using the equation below (Gimond, 2019):

$$I = \frac{\sum_i \sum_j w_{ij} z_i * z_j / S_o}{\sum_i z_i^2 / n}$$

where $w_{ij}$ is the spatial weight of location $i$ and $j$, $z_i$ is the observation at location $i$, $S_o = \sum_i \sum_j w_{ij}$ is the sum of all weights and $n$ is the number of observations. Besides the Moran's I statistic, a pseudo p-value is also calculated to provide information on the statistical significance of the spatial correlation.

For our analysis, taxi zones that share the same boundary are considered neighbors. All neighbors are weighted equally. Since Moran's I is often applied to cross-sectional data instead of time series, we aggregate the residuals at the following temporal resolutions:
- Full time series length
- 1/4 of the time series length
- Season – Spring, Summer, Fall, Winter
- Month

The residuals from Section 4 for TNC and Taxi are segmented into four equally sized sequential segments of 32 weeks (the first week was removed to get an integer number of weeks for each segment). Also, we aggregated all the series over the full length of time (2015-2017) to provide a reference. The results in Table 5 indicate that for all four segments for both modes (TNC and Taxi), the observed values of the global Moran's I statistic rejects the null hypothesis, indicating a significant degree of spatial correlation across the zones. Furthermore, it is worth noting that when we aggregate across the full length of the series, Moran's I fails to reject the null hypothesis of no spatial correlation for Taxi, but rejects the null hypothesis for TNC.

**Table 5 Global Moran's I – Segment Level Analysis**

|  | ARIMA-GARCH residuals | | | | ARIMA-GARCH-MLR residuals | | | |
|---|---|---|---|---|---|---|---|---|
|  | TNC | | Taxi | | TNC | | Taxi | |
|  | Moran's I Statistic | p-value | Moran's I Statistic | p-value | Moran's I Statistic | p-value | Moran's I Statistic | p-value |
| Segment 1 | 7.9680 | <0.001 | 4.9851 | <0.001 | 7.1904 | <0.001 | 3.9660 | <0.001 |
| Segment 2 | 3.2541 | <0.001 | 4.1059 | <0.001 | 2.5985 | 0.0047 | 3.1412 | <0.001 |
| Segment 3 | 9.1424 | <0.001 | 6.4023 | <0.001 | 6.9465 | <0.001 | 5.8728 | <0.001 |
| Segment 4 | 6.8790 | <0.001 | 2.5635 | 0.0052 | 1.8741 | 0.0305 | 1.2383 | 0.1078 |
| Full Series | 3.7491 | <0.001 | 1.5318 | 0.0628 | 1.7523 | 0.0399 | 1.1995 | 0.1152 |

Our next level of analysis was to see whether the spatial correlations are different between the four major seasons. We took all the time points corresponding to spring, summer, autumn, and winter, aggregated them for all the residual series and once again performed a global Moran's I test. Indeed, all four seasons exhibit spatial clustering as seen from the results in Table 6, with some slight differences in the values of the statistic between seasons.

Table 6 Global Moran's I -Season Level Analysis

| | ARIMA-GARCH residuals | | | | ARIMA-GARCH-MLR residuals | | | |
|---|---|---|---|---|---|---|---|---|
| | TNC | | Taxi | | TNC | | Taxi | |
| | Moran's I Statistic | p-value | Moran's I Statistic | p-value | Moran's I Statistic | p-value | Moran's I Statistic | p-value |
| Spring | 4.7672 | <0.001 | 6.7718 | <0.001 | 4.3515 | <0.001 | 6.5305 | <0.001 |
| Summer | 7.1486 | <0.001 | 3.7837 | 0.0001 | 7.0102 | <0.001 | 3.8943 | <0.001 |
| Autumn | 8.1210 | <0.001 | 5.8486 | <0.001 | 7.7366 | <0.001 | 5.8456 | <0.001 |
| Winter | 5.2768 | <0.001 | 0.4374 | 0.3309 | 5.2898 | <0.001 | 0.4810 | 0.3152 |

The results from a month-level analysis shown in Table 7 for also indicate spatial clustering for both TNC and Taxi usage.

Table 7 Global Moran's I -Monthly Level Analysis

| | ARIMA-GARCH residuals | | | | ARIMA-GARCH-MLR residuals | | | |
|---|---|---|---|---|---|---|---|---|
| | TNC | | Taxi | | TNC | | Taxi | |
| | Moran's I Statistic | p-value | Moran's I Statistic | p-value | Moran's I Statistic | p-value | Moran's I Statistic | p-value |
| January | 4.9744 | <0.001 | 5.2250 | <0.001 | 4.9848 | <0.001 | 5.1341 | <0.001 |
| February | 9.7074 | <0.001 | 10.0815 | <0.001 | 9.2391 | <0.001 | 9.9803 | <0.001 |
| March | 9.9091 | <0.001 | 6.7233 | <0.001 | 9.8670 | <0.001 | 6.6852 | <0.001 |
| April | 2.1195 | 0.017 | 5.8554 | <0.001 | 2.2448 | 0.0124 | 5.8475 | <0.001 |
| May | 7.3713 | <0.001 | 4.6212 | <0.001 | 6.9221 | <0.001 | 4.7753 | <0.001 |
| June | 7.5752 | <0.001 | 3.5595 | <0.001 | 8.1056 | <0.001 | 3.6554 | <0.001 |
| July | 5.7473 | <0.001 | 2.6634 | 0.004 | 5.4489 | <0.001 | 2.6596 | 0.0039 |
| August | 4.0230 | <0.001 | 5.0354 | <0.001 | 4.0695 | <0.001 | 5.0425 | <0.001 |
| September | 10.0105 | <0.001 | 3.7446 | <0.001 | 9.8920 | <0.001 | 3.7427 | <0.001 |
| October | 6.1406 | <0.001 | 4.6914 | <0.001 | 5.8578 | <0.001 | 4.6852 | <0.001 |
| November | 7.3623 | <0.001 | 3.2865 | <0.001 | 7.5311 | <0.001 | 3.2887 | <0.001 |
| December | 4.4517 | <0.001 | 5.6918 | <0.001 | 4.5969 | <0.001 | 5.6927 | <0.001 |

These results do not offer much insight into understanding spatial behavior across taxi zones. In Section 5.2, we explore the local Moran's I statistic.

**5.2 Local Moran's I**

While the global Moran's I explains the overall spatial patterns, the local Moran's I is a local indicator of spatial association of a zone with its neighborhoods. It is calculated by the following equations (Anselin, 1995):

$$I_i = \frac{z_i - \bar{Z}}{S_i^2} \sum_{j=1, j \neq i}^{n} w_{i,j}(z_j - \bar{Z})$$

$$S_i^2 = \frac{\sum_{j=1, j \neq i}^{n}(z_j - \bar{Z})^2}{n-1}$$

where $w_{ij}$ is the spatial weights of location $i$ and $j$, $z_i$ is the observation at location $i$ and $n$ is the number of observations.

We show results corresponding to local Moran's I to compare spatial associations across taxi zones in NYC for TNC and Taxi usage. Similar to the global Moran's I, local Moran's I values were also calculated at the same four temporal levels as we did in Section 5.2, i.e., full length (entire time series from 2015-2017), four segments (each of length 32 sequential weeks), four seasons (spring, summer, autumn, winter) and twelve months. Due to space limitations, we do not show the results from the analysis over the seasons and months here.

Figure 3 shows spatial heat maps based on values of the local Moran's I using all the data from 2015-2017; again, data here refers to residuals from fitting the ARIMA-GARCH model to each time series (plot in Figure 3 (a)) and residuals from fitting the ARIMA-GARCH-MLR model to each time series (plot in Figure 3 (b)). In each case, the plot on the left shows the local Moran's I statistics for TNC usage while the plot on the right shows these values for Taxi usage.

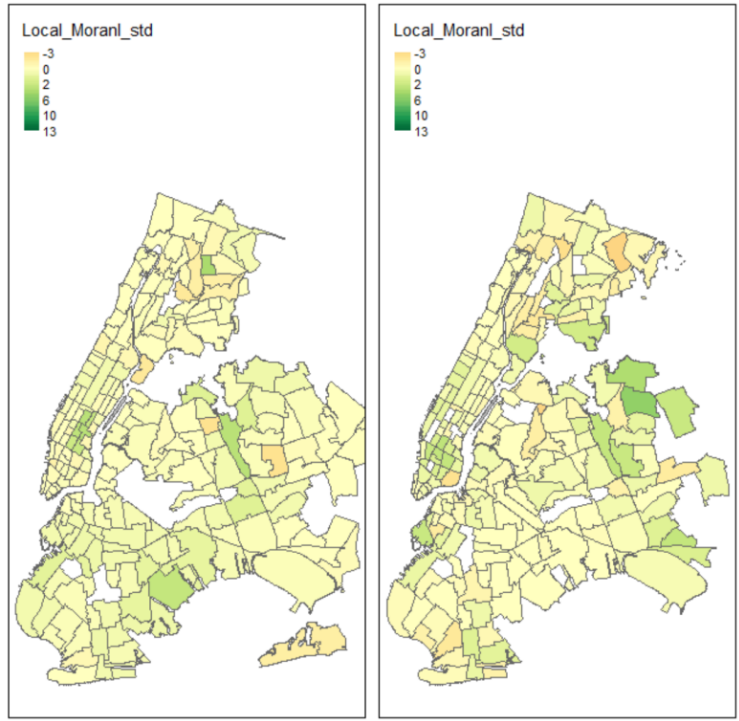

(a) ARIMA-GARCH Residuals

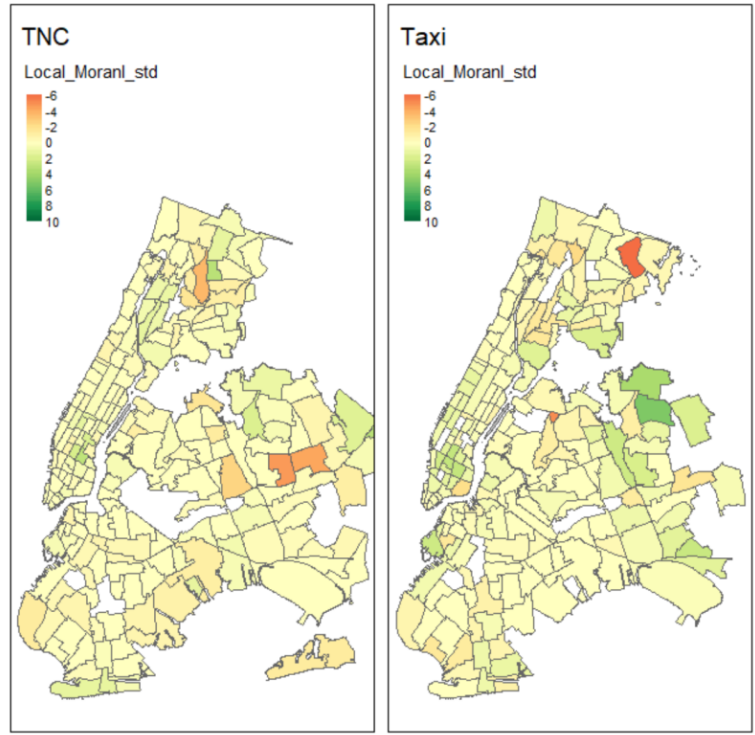

(b) ARIMA-GARCH-MLR Residuals

**Figure 3. Local Moran's I– Full Time Length (2015-2017)**

We compared the plots in (a) to the plots in (b) to see whether there are sizable spatial correlations between taxi zones before and after accounting for demographics and land use. We also compared the strength of the spatial correlations between TNC usage (left) and Taxi usage (right). In both plots (a) and (b), the patterns of Local Moran's I of TNC and Taxi are different.

In plot (a), we observe that three taxi zones (Whitestone, Murray Hill-Queens and Bayside) in Queens all have positive Local Moran's I values. This implies that these three zones have similar patterns in taxi ridership in their neighborhoods. While for TNC, the Moran's I values for these three zones are close to zero, which indicates no strong spatial correlation in TNC ridership in these zones. Flushing Meadows-Corona Park, Queens has a positive Moran's I for both TNC and Taxi, implying that both Taxi and TNC ridership patterns in this zone are similar to its neighborhoods. In Bronx, the taxi zone Bronxdale has a positive Moran' I for TNC ridership. Soundview/Castle Hill and Mott Haven/Port Morris all have a positive Moran' I for Taxi ridership. The Moran's I for other zones are close to zero or negative. In Manhattan, Gramercy, Midtown Center, Midtown South and Murray Hill all have positive Local Moran's I for TNC residuals. Flatiron, Hudson Square, Greenwich Village North, Greenwich Village South, Little Italy/NoLiTa and Union Square all have positive Local Moran's I for Taxi residuals. In Brooklyn, the taxi zone Canarsie has a positive Local Moran's I for TNC residuals, its Local Moran's I for taxi residuals is close to zero.

We also calculated the Local Moran's I statistics on the ARIMA-GARCH-MLR residuals (plot (b)) to see whether socioeconomic characteristics impact spatial correlations. It is interesting to see that plots (a) and (b) show clear differences. For example, the local Moran's I from the ARIMA-GARCH residuals for TNC in Flushing Meadows-Corona Park, Queens was positive, but changes to close a value close to zero after removing socioeconomic effects. Since Flushing Meadows-Corona Park is a large public park in NYC, its demographic characteristics are very different from its neighbors. The reason that this zone has a different TNC ridership pattern is likely to be caused by its different demographics.

Figure 4 shows the maps of the spatial correlations when the data is blocked into four segments each containing 32 weeks. We did this experiment to see whether the spatial dependence changed over time. For conciseness, we only show plots constructed from the local Moran's I values for the ARIMA-GARCH-MLR residuals.

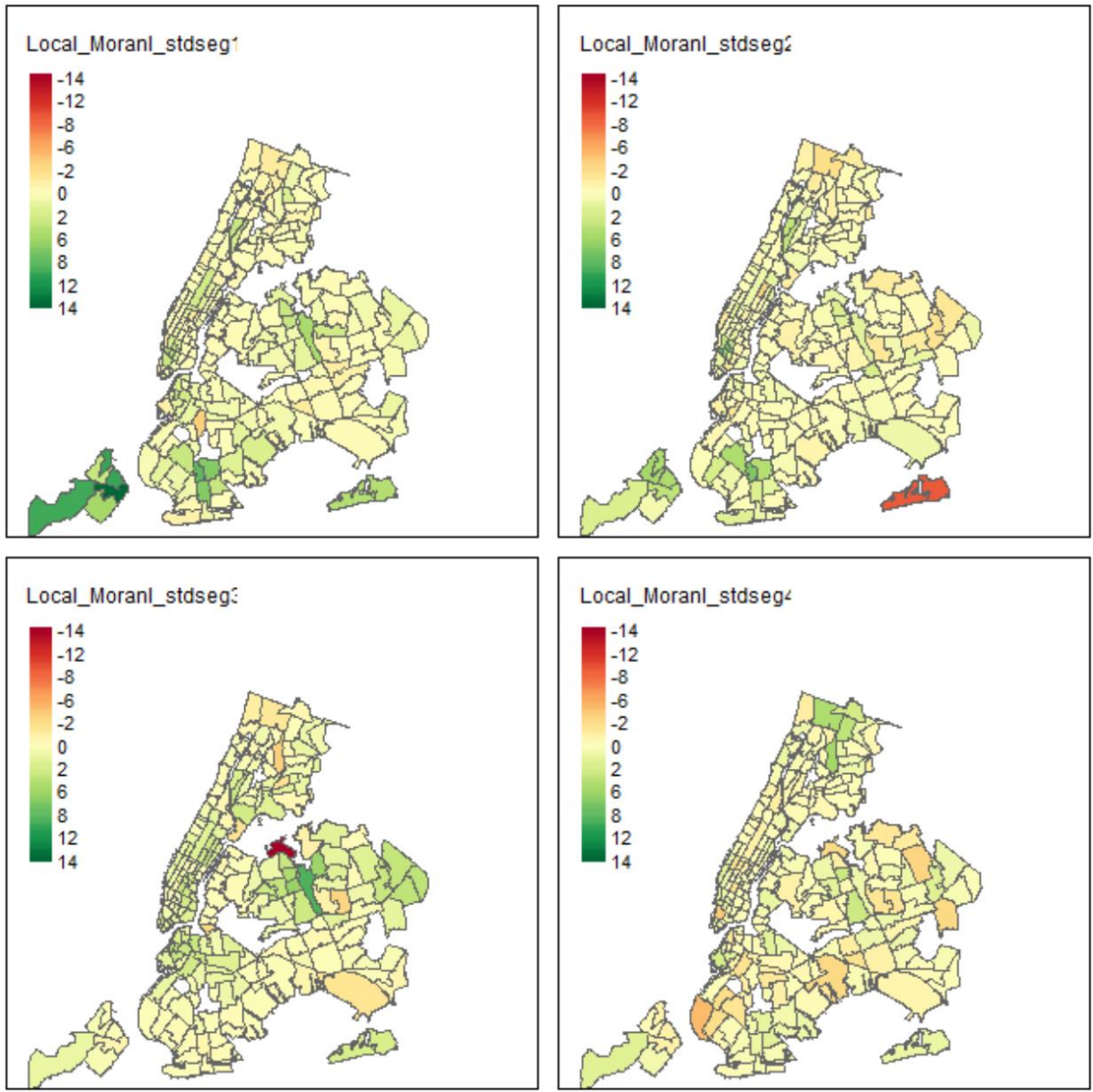

(a) TNC ARIMA-GARCH-MLR Residuals

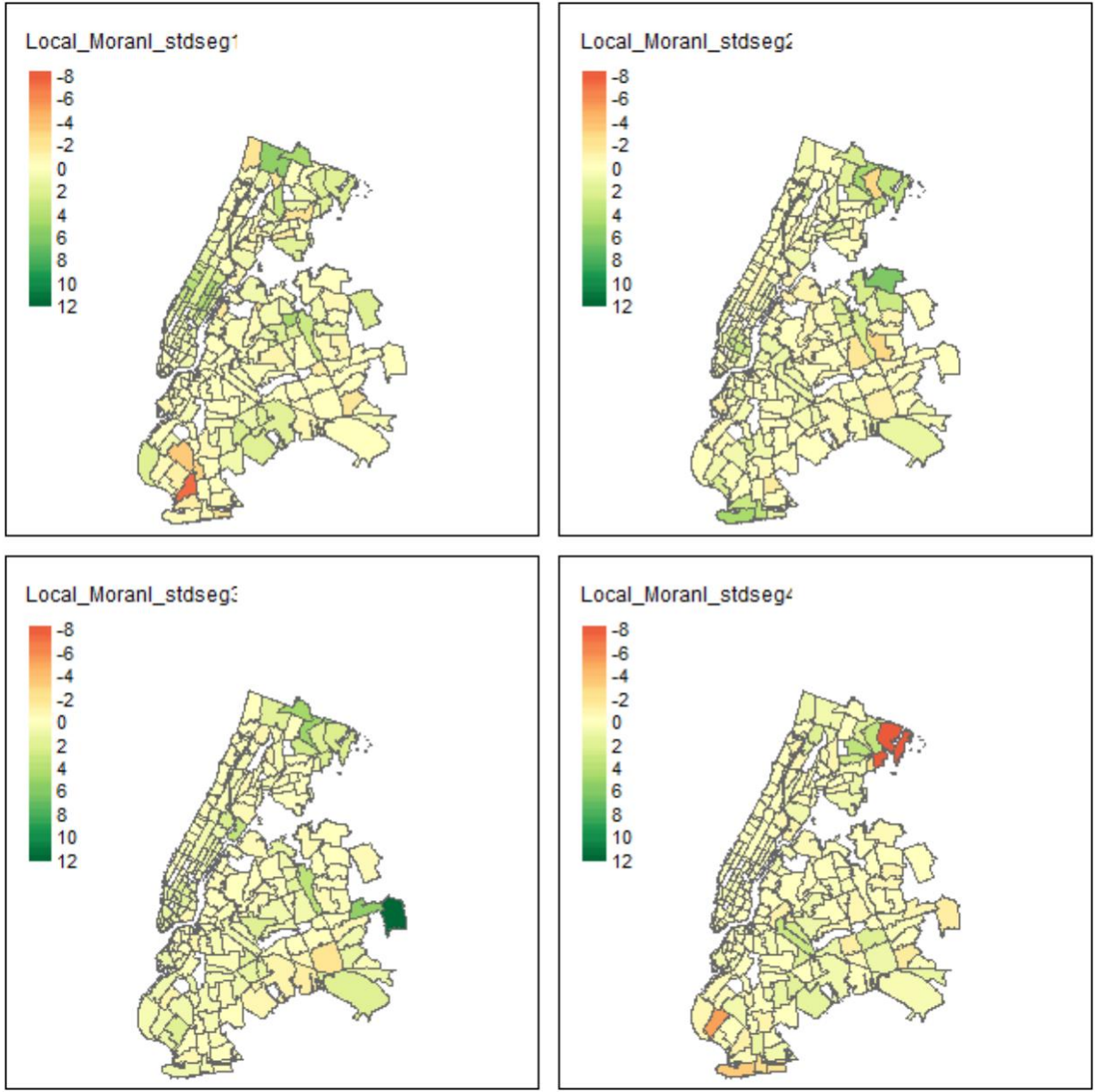

(a) Taxi ARIMA-GARCH-MLR Residuals

**Figure 4. Local Moran's I– Segment Level**

Figure 4(a) shows the Local Moran's I of TNC residuals over the four time segments and clearly changes over time. For example, taxi zones Hammels/Arverne and Far Rockaway in Queens both have a positive value of Local Moran's I in segments 1, 3 and 4, while their Moran's I values become negative in segment 2. This implies that the TNC ridership of these two zones have similar patterns as their neighborhoods in segments 1, 3 and 4, which is completely different from their neighborhoods in segment 2. Similarly, for LaGuardia Airport in Queens, the Local Moran's I is positive in segments 1 and 2, but becomes negative in segments 3 and 4. Figure 4(b) shows the Local Moran's I plot of Taxi residuals over four segments and shows that the Local Moran's I changes between the four segments. For Pelham Bay Park in Bronx, it is positive in segments 1, 2 and 3, becoming negative in segment 4. On the contrary, for Bensonhurst East in Brooklyn, it is negative in segment 1, changing to near zero in segments 2, 3 and 4. Although due to space constraints, we do not show these results, similar maps to compare the spatial associations between the four seasons or between the months of a year can be easily obtained. This analysis enables us to better understand the variations in spatial association over zones and across different periods.

## 6. Discussion and Summary

The primary goal of this analysis is to explore relationships between TNC and taxi modes at the taxi zone level in New York City. A temporal modeling of ridesourcing data using ARIMA-GARCH models have been first applied to remove temporal correlation in ridership data of two modes. It is followed by further removing demographic and land use characteristics from the residuals of ARIMA-GARCH models using an MLR model. Global and local Moran's I have been calculate using the resulting residuals from the models to understand the spatial associations of ridership of two modes at different temporal levels including: full length of series, four segments (32 weeks), seasonal and monthly level. Both global and local Moran's I indicate significant spatial clustering pattern of ridership of two modes. The plots of Local Moran's I show that the spatial patterns of TNC and Taxi ridership are different. For example, in Queens, three taxi zones - Whitestone, Murray Hill-Queens and Bayside have strong positive spatial correlations for Taxi ridership, while they do not show any strong spatial correlation for TNC ridership. The results also suggest that the spatial associations vary across different time periods even after removing temporally effects, suggesting that perhaps a dynamic model with time-varying coefficients or a may be helpful. The overall takeaway from this analysis is that predictive models for ridesourcing usage must consider spatial, temporal and socio-demographic factors.

## Acknowledgements

The authors are grateful to the Center for Advanced Multimodal Mobility Solutions and Education Year 3 funding for supporting this research and to Raymond Gerte for his help with data preparation.